\documentclass[journal]{IEEEtran}

\usepackage{booktabs}
\usepackage{bm}
\usepackage{graphicx}
\usepackage[cmex10]{amsmath}
\usepackage{cite}
\usepackage{amssymb}
\usepackage{subfigure}
\usepackage{extarrows}
\usepackage[letterspace = -2]{microtype}
\usepackage{algorithm}
\usepackage{algpseudocode}
\usepackage{float}
\usepackage{upgreek}
\usepackage{color}
\usepackage{makecell}

\usepackage[absolute,showboxes]{textpos}

\newfloat{algorithm}{t}{lop}

\begin{document}

\title{Distributed Multi-Objective Dynamic Offloading Scheduling for Air-Ground Cooperative MEC}

\author{Yang~Huang, Miaomiao~Dong, Yijie~Mao, Wenqiang~Liu, and Zhen~Gao
\thanks{Copyright \copyright  2015 IEEE. Personal use of this material is permitted. However, permission to use this material for any other purposes must be obtained from the IEEE by sending a request to pubs-permissions@ieee.org.}
\thanks{This work was partially supported by the National Natural Science Foundation of China under Grant U2001210, 62211540396, 61901216, and the Key R\&D Plan of Jiangsu Province under Grant BE2021013-4. (Corresponding author: Yijie Mao)}
\thanks{Y. Huang, M. Dong and W. Liu are with College of Electronic and Information Engineering, Nanjing University of Aeronautics and Astronautics, Nanjing, 210016, China (e-mail:\{yang.huang.ceie, dongmiaomiao, sx2304085\}@nuaa.edu.cn).}
\thanks{Y. Mao is with School of Information Science and Technology, ShanghaiTech University, Shanghai, 201210, China (e-mail:maoyj@shanghaitech.edu.cn).}
\thanks{Z. Gao is with State Key Laboratory of CNS/ATM, Beijing Institute of Technology, Beijing 100081, China; MIIT Key Laboratory of Complex-field Intelligent Sensing, Beijing Institute of Technology, Beijing 100081, China; Yangtze Delta Region Academy of Beijing Institute of Technology (Jiaxing), Jiaxing 314019, China; Advanced Technology Research Institute of Beijing Institute of Technology (Jinan), Jinan 250307, China; and Advanced Research Institute of Multidisciplinary Science, Beijing Institute of Technology, Beijing 100081, China
(e-mail: gaozhen16@bit.edu.cn).}
}

\maketitle

\begin{abstract}
Utilizing unmanned aerial vehicles (UAVs) with edge server to assist terrestrial mobile edge computing (MEC) has attracted tremendous attention.
Nevertheless, state-of-the-art schemes based on deterministic optimizations or single-objective reinforcement learning (RL) cannot reduce the backlog of task bits and simultaneously improve energy efficiency in highly dynamic network environments, where the design problem amounts to a sequential decision-making problem.
In order to address the aforementioned problems, as well as the curses of dimensionality introduced by the growing number of terrestrial terrestrial
users, this paper proposes a distributed multi-objective (MO) dynamic trajectory planning and offloading scheduling scheme, integrated with MORL and the kernel method. The design of n-step return is also applied to average fluctuations in the backlog.
Numerical results reveal that the n-step return can benefit the proposed kernel-based approach, achieving significant improvement in the long-term average backlog performance, compared to the conventional 1-step return design.
Due to such design and the kernel-based neural network, to which decision-making features can be continuously added, the kernel-based approach can outperform the approach based on fully-connected deep neural network, yielding improvement in energy consumption and the backlog performance, as well as a significant reduction in decision-making and online learning time.
\end{abstract}


\section{Introduction}
\label{INTRODUCTION}
Along with the widespread application of 5G technologies, various intelligent applications have emerged and gained widespread use in Internet of Things (IoT) devices \cite{LiuYG19-1,KGWGS20}.
However, stringent constraints on computing capability and power supply make IoT devices unable to cater to scenarios of computation-intensive and latency-critical services.
As a promising solution, mobile edge computing (MEC) enables IoT user equipment (UE) to offload computation tasks to edge servers. However, it is still difficult for MEC servers fixed at terrestrial base stations (BSs) to handle the scenario where existing infrastructures cannot satisfy unexpected increases in the demands of computation task processing.

Given the ability of high maneuverability, flexibility and employing line-of-sight (LoS) channels, MEC assisted with unmanned aerial vehicle (UAV) can be a potential countermeasure.
State of the art mainly focuses on scenarios where UEs can choose to perform computational tasks locally or offload to UAVs \cite{YGGG20,ZXLYX20,HWYZ19}.
Unfortunately, these offloading scheduling approaches are inapplicable to scenarios where UEs can offload task bits to the BS.
Besides, the approaches derived from deterministic optimizations and assumptions are inapplicable to the scenario where the channel gains and the statistical characteristics of producing the computational tasks are unknown to the network \cite{ZXLYX20}.
In practice, due to the continuous task production in a whole timeslot and the non-negligible time of signaling and data preparation for transmission, decision-making on offloading scheduling in a timeslot has no knowledge of the number of task bits produced in the timeslot, and the decisions for processing/offloading these task bits can only be executed in the next timeslot \cite{DPS16, Wang2022DynamicAC}. This boils down to a sequential decision-making problem.

\begin{figure}
\centering
\includegraphics[width = 3.4in]{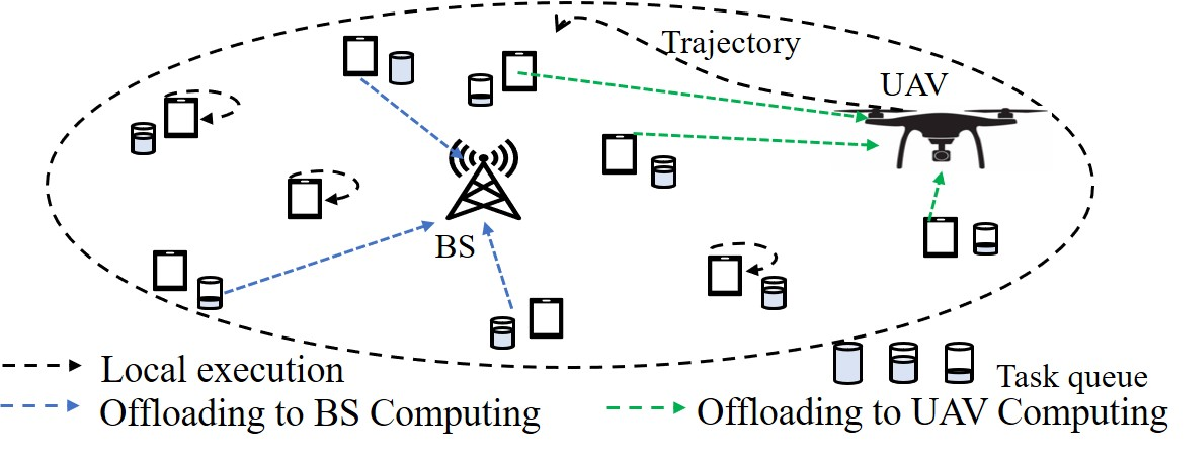}
\caption{Air-ground collaborative MEC with a UAV and a BS.}
\label{FigSystemModel}
\end{figure}
In order to handle the aforementioned issues, we focus on the scenario of air-ground collaborative MEC\cite{Wang2022DynamicAC} (as shown in Fig. \ref{FigSystemModel}), where a BS can be assisted by an edge server deployed at a UAV and reinforcement learning (RL) can be exploited to solve the sequential decision-making problem.
In contrast to the conventional single-objective optimization \cite{Wang2022DynamicAC}, in order to balance energy consumption and task backlog minimization, a novel multi-objective RL (MORL) approach is proposed to jointly optimize the trajectory planning and offloading scheduling policies.
In order to address the curses of dimensionality caused by the growing number of UEs, a distributed structure, where the overall network and UAV state is shared by all the agents while decision-making is performed at each agent, is integrated with the kernel method.
Numerical results demonstrate that the kernel-based approach with $n$-step return, which averages fluctuations in backlog of task bits, can achieve lower long-term average backlog than that with 1-step return.
Moreover, due to the $n$-step return and the kernel-based neural networks, where new features can be continuously learned and added, the kernel-based approach can significantly outperform the approach based on the deep neural network (DNN), in terms of the backlog performance and the average time consumption of decision-making and online learning.
It is also shown that the air-ground MEC can benefit from the online trajectory planning, fully utilizing air-ground channels and the UAV-mounted edge server.

\textit{Organization}: Section \ref{SecSystemModel} discusses the system model and formulate the problem.
Section \ref{SecJointTrajectOffloadAlgo} proposes the distributed multi-objective dynamic trajectory planning and offloading scheduling scheme.
Numerical results are analyzed in Section \ref{SecSimResults}.
Conclusions are drawn in Section \ref{SecConclusions}.

\textit{Notations}: Matrices and vectors are respectively in bold capital and bold lower cases; $(\cdot)^T$, $||\cdot||$ and $|\cdot|$ represent the transpose, $l_2$-norm, absolute value, respectively; cat($\mathbf{a}, \mathbf{b}$) concatenates $\mathbf{b}$ vertically to the end of $\mathbf{a}$.

\section{System Model and Problem Formulation}
\label{SecSystemModel}
\subsection{Network \& Communications}
%\subsection{Modeling of the Air-Ground Cooperative MEC}
\label{Network}
Without loss of generality, the studied system, as shown in Fig. \ref{FigSystemModel}, consists of a UAV flying at a constant altitude $H$, a BS and $M$ fixed terrestrial UEs, where a UE can offload computational task bits to the edge server deployed at the BS or the UAV-mounted edge server.

In terms of the terrestrial channel, due to the non-line-of-sight (NLoS) small-scale fading, we assume a block-fading channel model \cite{LCL2019}, where channel gains remain constant within a timeslot but vary across timeslots.
Therefore, in timeslot $t$, the channel power gain with respect to (w.r.t.) the channel from UE $m$ to the BS can be designated as $| h_{\text{BS},m,t} |^2  = |\Gamma_{\text{BS},m}|^2 |h_{\text{BS},0}|^2 d_{\text{BS},m}^{-\beta}$. The variables $\Gamma_{\text{BS},m}$, $h_{\text{BS},0}$, $d_{\text{BS},m}$ and $\beta$ represent the corresponding small-scale fading, the large-scale fading w.r.t. the terrestrial channel at a reference distance of 1\,m, the distance between the BS and UE $m$, and the pathloss exponent, respectively.

The ground-to-air channel can be characterized by the probabilistic LoS channel model \cite{AKL14}.
Therefore, given the position $\mathbf{q}_{\text{UAV},t}\!=\![x_{t},y_{t},H]$ of the UAV in a certain timeslot $t$, the LoS probability can be obtained as $P_\text{LoS} =  ({1 + a\exp{(-b(\arctan{({{H}/{r_{m,\text{UAV},t}}}) - a}))}})^{-1}$, where $a$ and $b$ represent constant modeling parameters; $r_{m,\text{UAV},t}$ represents the horizontal distance between UE $m \in \{1, \ldots, M\}$ and the UAV in timeslot $t$.
Let $h_{\text{UAV},0}$, $d_{\text{UAV},m,t}$ and $\Gamma_{\text{UAV},m}$ represent the large-scale fading w.r.t. the ground-to-air channel at a reference distance of 1\,m, the distance between UE $m$ and the UAV in timeslot $t$, and the small-scale fading w.r.t. the channel between UE $m$ and the UAV, respectively.
The channel power gain between UE $m$ and the UAV can be obtained as $|h_{\text{UAV},m,t}|^2 = |h_{\text{UAV},0}|^2 d_{\text{UAV},m,t}^{-2}$ with probability $P_\text{LoS}$; otherwise, $|h_{\text{UAV},m,t}|^2 = |\Gamma_{\text{UAV},m}|^2 |h_{\text{UAV},m,t,0}|^2 d_{\text{UAV},m,t}^{-\beta}$.
%\begingroup\makeatletter\def\f@size{9}\check@mathfonts
%\def\maketag@@@#1{\hbox{\m@th\small\normalfont#1}}%
%\begin{equation}
%|h_{\text{UAV},m,t}|^2=
%	\begin{cases}
%	|h_{\text{UAV},0}|^2 d_{\text{UAV},m,t}^{-2}\,,   \quad\text{with probability}\,\, P_\text{LoS}\\
%    \Gamma_{\text{UAV},m} |h_{\text{UAV},m,t,0}|^2 d_{\text{UAV},m,t}^{-\beta}, \quad \text{otherwise}\\
%	\end{cases}
%\end{equation}
%\endgroup

In each timeslot $t$, for each UE, the offloading scheduling options, including computing locally at a UE and offloading tasks to the UAV or the BS, are mutually exclusive.
The duration of offloading or/and executing tasks can be designated as $\tau$.
Assuming adequate number of frequency-domain channels, the UEs' offloading transmissions do not interfere with each other, and the computation results can be returned to the UEs via dedicated frequency-domain channels.
Therefore, the achievable rate at BS or the UAV in timeslot $t$ can be obtained as $R_{X,m,t}\! = \!B\log_{2}{(1\!+\!{{\left\vert h_{X,m,t} \right\vert}^2 P_m}/{\sigma_{n}^2})}$, where $B$, $P_m$ and $\sigma_n^2$ respectively represent the bandwidth, transmit power at UE $m$ and the average noise power. The value of the subscript $X$ depends on the edge server being deployed at the UAV or BS. In the former case, $X=\text{UAV}$; otherwise, $X=\text{BS}$.
Assume that a packet consists of a fraction of $\delta_b$ bits, the number $L_{m, X,t}$ of task bits that can be delivered during the duration of $\tau$ can be obtained as $L_{m, X,t} = \delta_b \lfloor R_{X,m,t}\cdot\tau/\delta_b \rfloor $.

\subsection{Computational Task Production and Processing}
Each UE continuously produces computational tasks over timeslots, and the statistical characteristics of task production is unknown\cite{ZXLYX20}.
Due to the overhead of signaling and data preparation \cite{9209079}, the task bits produced at each UE in an arbitrary timeslot $t$ can only be processed locally at the UE or offloaded in future timeslots.
%These $\Delta L_{m,t-1}$ task bits can only be
%The number of task bits produced by $m$th UE during timeslot $t-1$ can be designated as $\Delta L_{m,t-1}$.
%Hence, the produced task $L_{m,t-1}$ will be accumulated directly on the UE's local queue.

The CPU cycle frequencies of a certain UE, the edge server deployed at the BS and the UAV-mounted server are designated as $f_{\text{UE}}$, $f_{\text{BS}}$ and $f_{\text{UAV}}$, respectively.
The effective switched capacitance \cite{HWYZ19} w.r.t. the UE, the BS and the UAV are denoted by $\kappa_\text{UE}$, $\kappa_\text{BS}$ and $\kappa_\text{UAV}$, respectively.
The processing density \cite{zhang2018stochastic} is designated as $c$.
Task bits to be processed are buffered in queues, and the bits are processed following the first-in-first-out (FIFO) rule.
A variable $\alpha_{m,t, P}\!=\!\{0,1\} \, \text{for} \, P\in\{\text{UE,UAV,BS}\}$ is utilized to indicate the offloading action, including local processing, offloading task bits to the UAV, and offloading task bits to the BS, e.g. if UE $m$ performs local processing, $\alpha_{m,t,\text{UE}}\!=\!1$; otherwise, $\alpha_{m,t,\text{UE}}\!=\!0$.

In the presence of UE $m$ performing local processing in timeslot $t$, the exact time $t_{\text{cp},m,t}$ (which may not be equal to $\tau$) consumed for processing buffered task bits can be obtained, by following the modeling in \cite{Wang2022DynamicAC}.
Specifically, prior to formulating $t_{\text{cp},m,t}$ (for each timeslot), we can designate the number of task bits produced by UE $m$ during timeslot $t-1$ as $L_{m,t-1}$. Besides, the number of unprocessed task bits, which are observed at the end of timeslot $t-1$ but produced before timeslot $t-1$, can be designated as $D_{m, t-1}$.
When UE $m$ processing the buffered task bits in timeslot $t$, these $D_{m,t-1}$ bits take priority and be processed first, following the FIFO rule.
Meanwhile, the $L_{m,t-1}$ task bits generated in timeslot $t-1$ are accumulated at the end of the queue.
Then, we first evaluate the potential time $\hat{t}_{\text{cp},m,t}$ (which may be higher than $\tau$) for processing all the buffered bits. That is, $\hat{t}_{\text{cp},m,t}=c\cdot (L_{m,t-1}+D_{m,t-1})/f_{\text{UE}}$.
If $\hat{t}_{\text{cp},m,t}\le \tau$, all the $L_{m,t-1}+D_{m,t-1}$ bits can be processed within $\tau$, such that $t_{\text{cp},m,t}=c\cdot (L_{m,t-1}+D_{m,t-1})/f_{\text{UE}}$ and $D_{m,t}=0$;
otherwise, UE $m$ cannot complete processing all the task bits buffered in the previous timeslots, such that $t_{\text{cp},m,t}=\tau$ and $D_{m,t}=(L_{m,t-1}+D_{m,t-1})-f_{\text{UE}}\cdot\tau/c$.
Then, the energy consumed by local processing can be evaluated by $E_{\text{UE},m,t}^\text{cp} \! = \! \kappa_\text{UE} f_\text{UE}^3 t_{\text{cp},m,t}$.

In the presence of UE $m$ perform offloading in timeslot $t$, $L_{m,X,t}$ task bits (encapsulated in packets) can be offloaded to the edge server deployed at the UAV or BS during the duration of $\tau$.
The energy and the time consumed for transmission can be respectively achieved by $E_{X,m,t}^\text{trans} = {P_{m} L_{m,X,t}}/{R_{X,m,t}}$ and $t_{X,m,\text{trans}}\!=\!L_{m,X,t}/R_{X,m,t}$.
Besides receiving the offloaded bits, the edge server can simultaneously process tasks which are buffered in its task queue during previous timeslots.
By designating the number of the unprocessed task bits (which are observed at the end of timeslot $t-1$) as $D_{X,t-1}$, the potential time for processing the previously buffered task bits can be obtained as $t_{X, \text{pre}} = D_{X,t-1} \cdot c/f_X$.
If $t_{X,\text{pre}} \leq \tau$, all the $D_{X,t-1}$ bits can be processed before the end of offloading transmission.
It means that the residual time $t_{m,\text{res}} = \tau-t_{X,\text{pre}}$ can be exploited by the edge sever to process the received task bits (which are offloaded by the UE).
If the potential total processing time $L_{m,X,t} \cdot c/f_{X} > t_{m,\text{res}}$, the exact time consumed for processing is equal to $t_{\text{cp},X,m,t} = \tau$, and the number of unprocessed bits becomes $D_{X,t} = D_{X,t-1} + L_{m,X,t} - f_{X}\cdot\tau/c$;
otherwise, $t_{\text{cp},X,m,t} = t_{X,\text{pre}} + L_{m,X,t} \cdot c/f_{X}$, and $D_{X,t} = 0$.
On the other hand, in the case of $t_{X,\text{pre}} \geq \tau$, the edge server cannot complete the process of the task bits buffered in the previous timeslots. Therefore, the exact time consumed for processing task bits can be obtained as $t_{\text{cp},X,m,t} = \tau$, and the number of unprocessed bits reserved in the queue can be updated as $D_{X,t} = D_{X,t-1} + L_{m,X,t} - f_{X} \cdot \tau/c$.
Meanwhile, the energy consumed for processing task bits can be obtained as $E_{X,m,t}^\text{cp} = \kappa_{X}f_{X}^3 t_{\text{cp},X,m,t}$.

Given the above, the total energy consumed for transmission and processing can be achieved as $E_t = \sum_{m=1}^{M} (E_{\text{UAV},m,t}^\text{trans} \cdot \alpha_{m,t,\text{UAV}} + E_{\text{BS},m,t}^\text{trans} \cdot \alpha_{m,t,\text{BS}} + E_{\text{UE},m,t}^\text{cp} \cdot \alpha_{m,t,\text{UE}} + E_{X,m,t}^\text{cp} \cdot  (1  - \alpha_{m,t,\text{UE}}))$.
The total backlog of task bits can be obtained as $D_t = \sum_{m=1}^{m=M} (D_{m,t})  + D_{\text{UAV},t} +  D_{\text{BS},t}$.

\subsection{Problem Formulation}
\label{SecMEMDPnProbFormulation}
Aiming at minimizing the expected long-term average energy consumption and backlog of task bits, this study addresses the problem of optimizing policies for joint offloading scheduling and UAV trajectory planning.
Intuitively, the decision-making on trajectory planning and offloading scheduling turn out to be a Markov decision process (MDP) with two distinct objectives \cite{Wang2022DynamicAC, huang2021dynamic}, where the immediate reward can be therefore formulated as a vector $\mathbf{r}_t \!=\! [e_t,d_t]^T$, where $e_t = - E_t$ and $d_t = - D_t$.
For such an MDP, the UAV/backlog state in timeslot $t$ can be formulated as $\mathbf{s}_t = [\mathbf{q}_{\text{UAV},t},d_{t-1}^\prime]^T \in \mathcal{S}$, where $d_{t-1}^\prime = - \log(D_{t-1})$ and $\mathcal{S}$ represents the state space.
The trajectory planning action w.r.t. the UAV, i.e. the direction the UAV decides to fly towards, can be designated as $\mathbf{a}_{t,0} \in \mathcal{A}_0$, where $\mathcal{A}_0$ represents the UAV's action space that collects $\mathbf{a}_{t,0}$.
The offloading scheduling action that could be selected by a certain UE $m \in \{1,2, \ldots, M\}$ in timeslot $t$ is formulated as $\mathbf{a}_{t,m} = [\alpha_{m,t,\text{UAV}}, \alpha_{m,t,\text{BS}}, \alpha_{m,t,\text{UE}}]^T \in \mathcal{A}_m$ , where $\mathcal{A}_m$ represents the offloading scheduling action space w.r.t. UE $m$.
By defining $\mathcal{A} = \bigcup_{m=0}^M \mathcal{A}_m$, the joint trajectory planning and offloading scheduling policy can be expressed as $\pi: \mathcal{S} \rightarrow \mathcal{A}$.

In order to improve the expected long-term average energy efficiency and backlog of task bits, the average rewards
%w.r.t. $e_{t}$ and $d_{t}$
can be respectively obtained as $\bar{E} = \lim_{T \to \infty} \sup \mathcal{E}\{\sum_{t=0}^{T-1}e_{t}\} /T $ and $\bar{D} =  \lim_{T \to \infty} \sup \mathcal{E} \{\sum_{t=0}^{T-1}d_{t}\}/T$.
By collecting the rewards within $\overline{\mathbf{r}}\!\triangleq \![\bar{E}, \bar{D}]^{T}$ and defining $\mathbf{w}_r \triangleq [w_e,w_d]^{T}$, the optimization of $\pi$ can be formulated as

%\begingroup\makeatletter\def\f@size{9}\check@mathfonts
%\def\maketag@@@#1{\hbox{\m@th\small\normalfont#1}}%
\begin{equation}
\label{ProbOptimPi}
\pi=\arg\max_{\pi}\{\mathbf{w}_r^{T}\mathbf{\bar{r}}\}.
\end{equation}
%We denote the negative value of the energy consumption and the negative value of the sum of task bits backlog respectively as $e_{t}\!=\!-E_t$ and $d_{t}\!=\!-D_t$. %in timeslot $t$.

\section{Distributed Multi-Objective Dynamic Trajectory Planning \& Offloading Scheduling}
\label{SecJointTrajectOffloadAlgo}
The unknown statistics of computational task production and terrestrial channels make dynamics of $P(\mathbf{s}_{t+1} = \mathbf{s}^\prime|\mathbf{s}_{t} = \mathbf{s}, \mathbf{a}_{t,m} = \mathbf{a}_{m})$ $\forall \mathbf{s},\mathbf{s}^\prime \in \mathcal{S}$, $\forall \mathbf{a}_{m},\mathbf{a}_{m} \in \mathcal{A}_{m}$ unknown for solving problem (\ref{ProbOptimPi}). Intuitively, such a problem with multiple objectives can be addressed by MORL \cite{huang2021dynamic, LXH18}.

\subsection{Kernel-Based Approach with $n$-step Return}
\label{SubSecKernelApproach}
A centralized decision-making for all the UEs and the UAV (i.e. the agents) can suffer from the curses of dimensionality \cite{1998Reinforcement}.
As a countermeasure, we propose a distributed MORL, where each agent can select the offloading action or flight direction within its own action space $\mathcal{A}_m$.

The agents share the observation of $\tilde{\mathbf{s}}_{t}$, which is a quantized version of $\mathbf{s}_t$, i.e. $\tilde{\mathbf{s}}_{t} \!=\!  [{\tilde{\mathbf{q}}_{\text{UAV},t}} , \tilde{d}_{t-1}^\prime]^T \in \tilde{\mathcal{S}}$.
Note that elements in the set of $\tilde{\mathcal{S}}$ are not predefined but added into the set during online learning or offline training over timeslots.
Specifically, given thresholds $\mu_q$ and $\mu_d$, if a newly observed $\tilde{\mathbf{s}}_{t}$ in timeslot $t$ satisfies $\|\tilde{\mathbf{q}}_{\text{UAV},t} \! - \! \tilde{\mathbf{q}}_\text{UAV}\| \!  > \! \mu_q$ or $|\tilde{d}_{t-1}^\prime \! - \! \tilde{d} | \! > \! \mu_d$ for all $\tilde{\mathbf{s}} \! = \! [\tilde{\mathbf{q}}_\text{UAV}, \tilde{d}]^T \! \in \! \tilde{\mathcal{S}}$,
$\tilde{\mathbf{s}}_{t}$ is recognized as a new state and $\tilde{\mathcal{S}} \! = \! \tilde{\mathcal{S}} \! \cup \! \mathbf{\tilde{s}}_{t}$.
Then, in order to reduce the number of optimization variables w.r.t. policies $\pi_m\!: \mathcal{S}\! \rightarrow \! \mathcal{A}_m\, \forall m$ (and the elapsed time for training/inference), the action-values are approximated with linearly combined Gaussian kernels.
Therefore, given $\tilde{\mathbf{s}}\in \tilde{\mathcal{S}}$, the action-values w.r.t. maximizing $\bar{E}$ and $\bar{D}$ (i.e. minimizing the long-term average energy consumption and backlog of task bits) can be respectively expressed as $Q_{e,m}(\tilde{\mathbf{s}},\mathbf{a}_{m};\mathbf{w}_{e,m})\!=\!\mathbf{w}_{e,m}^{T}\mathbf{f}_{e,m,t}(\tilde{\mathbf{s}},\mathbf{a}_{m})$ and $Q_{d,m}(\tilde{\mathbf{s}},\mathbf{a}_{m};\mathbf{w}_{d,m})\!=\!\mathbf{w}_{d,m}^{T}\mathbf{f}_{d,m,t}(\tilde{\mathbf{s}},\mathbf{a}_{m})$, where $\mathbf{w}_{e,m}$ and $\mathbf{w}_{d,m}$ represent weight vectors to be optimized over iterations, i.e. the learning process;
$\!\mathbf{f}_{e,m,t}$ and $\!\mathbf{f}_{d,m,t}$ are kernel vectors with $N_{e,m,t}$ and $N_{d,m,t}$ entries, respectively.
Moreover, each entry of $\mathbf{f}_{e,m,t}$ can be written as $[\mathbf{f}_{e,m,t}]_{n}\!=\!f(\mathbf{x}_{e,m},\mathbf{\widehat{x}}_{e,m,n})\!=\! \phi(\mathbf{x}_{e,m})^{T}\phi(\mathbf{\widehat{x}}_{e,m,n}) \!=\!\exp(-\|\tilde{\mathbf{q}}_{\text{UAV}}\!-\!\mathbf{\widehat{q}}_{n} \|^2 /2\sigma_{s_1}^2)\exp(-\|\tilde{{d}^\prime}\!-\!{\widehat{d}}_n \|^2 /2\sigma_{s_2}^2)\exp(-\|\mathbf{a}_{m}\!-\!\mathbf{\widehat{a}}_{m,n} \|^2 /2\sigma_a^2)$ for $n\!=\!1, \ldots, N_{e,m,t}$, where $\mathbf{x}_{e,m} \triangleq [\tilde{\mathbf{s}}^T, \mathbf{a}_{m}]^T$, $\mathbf{\widehat{x}}_{e,m,n} \triangleq [\mathbf{\widehat{s}}_{n}^T, \mathbf{\widehat{a}}_{m,n}]^T$ and $\phi(\cdot)$ respectively represent a certain sample, the $n$\,th feature and the feature space mapping w.r.t. decision-making; $\sigma_{s_1}$, $\sigma_{s_2}$ and $\sigma_a$ respectively denote the characteristic length scales w.r.t. the feature vectors of $\mathbf{\widehat{q}}_{n}$, $\widehat{d}_n$ and $\mathbf{\widehat{a}}_{m,n}$.
Each entry of $\mathbf{f}_{d,m,t}$ is formulated in similar manner.
All features are collected in the sets of $\mathcal{D}_{e,m,t} \! \triangleq \! \{\mathbf{\widehat{x}}_{e,m,n}\}_{n=1}^{N_{e,m,t}} $ and $\mathcal{D}_{d,m,t} \! \triangleq \! \{\mathbf{\widehat{x}}_{d,m,n}\}_{n=1}^{N_{d,m,t}}$.

In order to select a trajectory planning/offloading scheduling action maximizing the weighted objective as shown in (\ref{ProbOptimPi}), a synthetic vector-valued function can be defined as  $\mathbf{q}_m(\tilde{\mathbf{s}}_{t}, \mathbf{a}_{t,m})\!=\! [Q_{e,m}(\tilde{\mathbf{s}}_{t}, \mathbf{a}_{t,m}),Q_{d,m}(\tilde{\mathbf{s}}_{t},\mathbf{a}_{t,m})]^{T}$ \cite{huang2021dynamic}.
Therefore, given $\tilde{\mathbf{s}}_{t}$, an optimized action can be obtained as
%\begingroup\makeatletter\def\f@size{9}\check@mathfonts
%\def\maketag@@@#1{\hbox{\m@th\small\normalfont#1}}%
\begin{equation}
\label{Select_action}
\mathbf{a}_{m}^\star = \arg \max_{\mathbf{a}_{m}\in\mathcal{A}_{m}} \left\{\mathbf{w}_r^T\mathbf{q}_{m}(\tilde{\mathbf{s}}_{t},\mathbf{a}_{t,m})\right\}.
\end{equation}
%\endgroup
A conventional $\epsilon$-greedy strategy for exploration and exploitation can yield a strictly suboptimal action $\mathbf{a}_{t,m}$ with a certain probability even if $\pi_m$ converges. Thus, an improved $\epsilon$-greedy strategy is developed, such that only actions that have not been visited before can be explored.
To this end, we first define $\mathbf{T}_{m}$ for each agent $m$ to indicate the visit state of state-action pair.
That is, if a pair of state $j$ and action $k$ has been visited, $[\mathbf{T}_m]_{j,k} = 1$; otherwise, $[\mathbf{T}_m]_{j,k} = 0$.
Note that in the presence of a certain $\tilde{\mathbf{s}}_{t}$ being recognized as a new state, $\mathbf{T}_m \!=\! \text{cat}( \mathbf{T}_m, \mathbf{0}_{1 \times | \mathcal{A}_m | } )$.
Given $\mathbf{T}_{m}$ and a certain $\tilde{\mathbf{s}}_{t} = \tilde{\mathbf{s}}_j$, with probability $\epsilon$, $\mathbf{a}_{t,m}$ is randomly selected from the set of actions w.r.t. $[\mathbf{T}_m]_{j,k}= 0$; otherwise, (\ref{Select_action}) is performed to obtain $\mathbf{a}_{t,m}^\ast$.

In order to address such RL with an average reward RL as shown in (\ref{ProbOptimPi}) and optimizing $\mathbf{w}_{e,m}$ and $\mathbf{w}_{d,m}$ over iterations, we integrate R-learning \cite{Mahadevan1996Average} with the semi-gradient method.
Moreover, as the immediate reward $d_t$ may periodically fluctuate even in the presence of a fixed $\pi_m$ \cite{Wang2022DynamicAC}, the $n$-step return method \cite{1998Reinforcement} is exploited.
To this end, we define $\mathbf{r}_{m,t:t+n}\!=\![e_{t:t+n},d_{t:t+n}]^T$ and $\mathbf{r}_{m,t:t+n}\!\leftarrow \!\mathbf{r}_{m,t+1}+\gamma_r\mathbf{r}_{m,t+2:t+n}$, where $\mathbf{r}_{m,t+1}$ and $\gamma_r$ respectively represent the reward in timeslot $t\!+\!1$ and the discount rate (also known as a discount factor).
Therefore, let $\alpha$ and $k_r$ denote the learning rates, $\mathbf{w}_{e,m}$, $\mathbf{w}_{d,m}$ and the estimated average reward $\mathbf{\bar{r}}_{m,t}=[\bar{E}_{t},\bar{D}_{t}]^T$ can be updated by performing
%\begingroup\makeatletter\def\f@size{9}\check@mathfonts
%\def\maketag@@@#1{\hbox{\m@th\small\normalfont#1}}%
\begin{IEEEeqnarray}{rl}
\mathbf{w}_{e,m} \! \leftarrow & \mathbf{w}_{e,m} \! + \! \alpha \big(e_{m,t:t+n} \! + \! \textcolor{blue}{\gamma_r} \!  \max_{\mathbf{a}_m} \{ \! \mathbf{w}_{e,m}^{T}\mathbf{f}_{e,m,t}(\tilde{\mathbf{s}}_{t\!+\!1},\mathbf{a}_{m})\} \nonumber \\
& -\bar{E}_{t}- \mathbf{w}_{e,m}^{T} \mathbf{f}_{e,m,t}(\tilde{\mathbf{s}}_{t},\mathbf{a}_{t,m}) \big) \mathbf{f}_{e,m,t}(\tilde{\mathbf{s}}_{t},\mathbf{a}_{t,m})\,, \label{UpdateQ_1} \\
\mathbf{w}_{d,m} \! \leftarrow & \mathbf{w}_{d,m} \! + \! \alpha(d_{m,t:t+n} \!+ \!\textcolor{blue}{\gamma_r} \! \max_{\mathbf{a}_m} \{ \! \mathbf{w}_{d,m}^{T}\mathbf{f}_{d,m,t}(\tilde{\mathbf{s}}_{t\!+\!1} \! ,\mathbf{a}_{m})\} \nonumber \\
& -\bar{D}_{t}-\mathbf{w}_{d,m}^{T}\mathbf{f}_{d,m,t}(\tilde{\mathbf{s}}_{t},\mathbf{a}_{t,m})) \mathbf{f}_{d,m,t}(\tilde{\mathbf{s}}_{t},\mathbf{a}_{t,m})\,, \label{UpdateQ_2} \\
\mathbf{\bar{r}}_{m,t+1}\! =  {}& \mathbf{\bar{r}}_{m,t}(1\! -\!k_{r}) \!+\! k_{r}(\mathbf{r}_{m,t:t+n} \!+\! \mathbf{q}_{m}(\tilde{\mathbf{s}}_{t+1},\mathbf{a}_{m}^\star) \nonumber \\
& {} -{} \mathbf{q}_{m}(\tilde{\mathbf{s}}_{t},\mathbf{a}_{t,m}))\,, \label{Requation}
\end{IEEEeqnarray}
%\endgroup
where $\mathbf{a}_{m}^\star$ can be obtained by performing (\ref{Select_action}). Moreover, (\ref{Requation}) is performed, only in the presence of $\mathbf{a}_{t,m}$ not being generated by exploration \cite{Mahadevan1996Average}.
\begin{algorithm}[t]
\caption{Kernel-Based Approach with $n$-step Return}\label{Algkernelbased}
%The Kernel-Based Approach
\begin{algorithmic}[1]
\State \textbf{Initialize:} For $m=0,\cdots,M$, initialize $\mathbf{s}_0$, $\tilde{\mathbf{s}}_0$, $\mathcal{S}$; set $\tilde{\mathcal{S}}=\tilde{\mathbf{s}}_0$
$\mathbf{T}_{m}=\mathbf{0}_{1\times| \mathcal{A}_m |}$ $\forall m$; set $t=1$; initialize $\mathbf{w}_{m,e}$ and $\mathbf{w}_{m,d}$.
\Repeat
 \State Each agent $m$ observes $\mathbf{s}_{t}$ and quantizes $\mathbf{s}_{t}$ , yielding $\tilde{\mathbf{s}}_{t}$;
  \If {$\tilde{\mathbf{s}}_{t}\notin{\mathcal{S}}$ }
  {\hfill $\rhd$ Check if $\tilde{\mathbf{s}}_{t}$ is a new state}
  \State $\mathcal{S}=\mathcal{S}\cup \tilde{\mathbf{s}}_{t}$ and  $\mathbf{T}_m = \text{cat}( \mathbf{T}_m, \mathbf{0}_{1 \times | \mathcal{A}_m | } )$;
  \EndIf
%\State In the timeslot $t$, agent-UAV finds the row index $j$ w.r.t. $\tilde{\mathbf{s}}_{t}$ in $\mathbf{T}_{0}$ and obtain $\mathcal{A}_{0,j}$=
%$\left\{ \mathbf{a}_{0}| \text{all}\;\mathbf{a}_{0}\in\mathcal{A}_{0}\text{for which}\;[{\mathbf{T}_{m}]_{j,k}=0} \right\}$;% select $\mathbf{a}_{t,0}$via %\ref{Selectaction}
%\State Generate a random number $\epsilon_x\sim\mathcal{U}(0,1)$
%    \If{$\epsilon_x<\epsilon$ }
%    {\hfill $\rhd$ Flight maneuver configuration for UAV}
%            \State Given $\tilde{\mathbf{s}}_{t}$, obtain $\mathbf{a}_{t,0}$ by randomly select an action $\mathbf{a}_{t,0}\! \in \! \mathcal{A}_{0,j}$;
%     \Else
%             \State Given $\tilde{\mathbf{s}}_{t}$, compute $\mathbf{a}_{t,0}$ by solving (\ref{Select_action});
%     \EndIf
%\State Set$[{\mathbf{T}_{0}]_{j,k}}$ corresponding to $(\tilde{\mathbf{s}}_{t},\mathbf{a}_{t,0})$ as $[{\mathbf{T}_{0}]_{j,k}=1}$;
\For {$m=0, \ldots, M$}
   {\hfill $\rhd$ Action configuration for agent}
\State Each agent $m$ finds the row index $j$ w.r.t. $\tilde{\mathbf{s}}_{t}$ in $\mathbf{T}_{m}$ and obtain $\mathcal{A}_{m,j}$=\!
$\left\{ \mathbf{a}_{m}| \text{all}\;\mathbf{a}_{m}\!\in\!\mathcal{A}_{m}\text{for which}\;[{\mathbf{T}_{m}]_{j,k}\!=\!0} \right\}$; %select $\mathbf{a}_{t,m}$ via
\State Generate a random number $\epsilon_x\sim\mathcal{U}(0,1)$;
    \If{$\epsilon_x<\epsilon$ }
            \State Given $\tilde{\mathbf{s}}_{t}$, randomly select an action $\mathbf{a}_{t,m}\! \in \! \mathcal{A}_{m,j}$;
            %obtain $\mathbf{a}_{t,m}$ by randomly select an action $\mathbf{a}_{t,m}\! \in \! \mathcal{A}_{m,j}$;
     \Else
             \State Given $\tilde{\mathbf{s}}_{t}$, compute $\mathbf{a}_{t,m}$ by solving (\ref{Select_action});
     \EndIf
\State Set $[{\mathbf{T}_{m}]_{j,k}}$ corresponding to $(\tilde{\mathbf{s}}_{t},\mathbf{a}_{m,t})$ as $[{\mathbf{T}_{m}]_{j,k}=1}$;
\EndFor
\State In timeslot $t$, for $m=0,\cdots,M$, each agent $m$ executes $\mathbf{a}_{t,m}$ and obtains reward $\mathbf{r}_{m,t+1}$;
\For {$m=0, \ldots, M$}
\State $\rho\!\leftarrow\!t\!-\!n\!+\!1$
   \If {$\rho\!\ge\!0$}
   {\hfill $\rhd$ Obtain reward for $N$ steps}
   \State $\mathbf{r}_{m,t:t+n}\!\leftarrow\!{\sum_{i=\rho+1}^{\text{min}\{\rho+n,T\}}}\gamma_r^{i-\rho-1}\mathbf{r}_{m,i}$;
   \EndIf
   %{\hfill $\rhd$ Obtain reward for $N$ steps}
\State Update $\mathbf{w}_{e,m}$ and $\mathbf{w}_{d,m}$ by performing (\ref{UpdateQ_1}) and (\ref{UpdateQ_2});
   \If {$\epsilon_x\ge\epsilon$}
   \State Update $\mathbf{\bar{r}}_{m,t+1}$ by performing (\ref{Requation});
   \EndIf
%\State Update the weight vectors $\mathbf{w}_{e,m}$ and $\mathbf{w}_{d,m}$ of the stored features in dictionary $\mathcal{D}_{m,e,t}$ and $\mathcal{D}_{m,d,t}$;\label{algorithm18}
\State Perform ALD test to update $\mathcal{D}_{d,m,t}$ and $\mathcal{D}_{e,m,t}$;
\EndFor
\State $t \leftarrow t+1$;\label{algorithm20}
\Until{\text{Stopping criteria}
\end{algorithmic}}
\end{algorithm}
New decision-making features can be added into $\mathcal{D}_{e,m,t}$ and $\mathcal{D}_{d,m,t}$ so as to improve the approximation of $Q_{e,m}(\tilde{\mathbf{s}}_{t}, \mathbf{a}_{t,m})$ and $Q_{d,m}(\tilde{\mathbf{s}}_{t},\mathbf{a}_{t,m})$, by performing the approximate linear dependence (ALD) \cite{EYMSMR2004} test.
In terms of the update of $\mathcal{D}_{e,m,t}$, given a threshold $\mu_0$, if $\delta_{0,t} = \min_{\lambda_{n}\forall n}\| \sum_{n=1}^{N_{e,m,t}} \lambda_n\phi_n(\mathbf{\widehat{x}}_{e,m,n})\!-\!\phi(\mathbf{x}_{e,m,t}) \|^2 \leq \mu_0$, it means that $\phi(\mathbf{x}_{e,m,t})$ can be approximated by $\{ \phi(\widehat{\mathbf{x}}_{e,m,n}) \}_{n=1}^{N_{e,m,t}}$; otherwise, $\mathcal{D}_{e,m,t+1} \!=\!\mathcal{D}_{m,e,t}\cup\mathbf{x}_{e,m,t}$. The update of $\mathcal{D}_{d,m,t}$ is similar to that of $\mathcal{D}_{e,m,t}$. The above proposed algorithm is then referred to as the kernel-based approach and is summarized in Algorithm \ref{Algkernelbased}.

\subsection{DNN-Based Approach}
\label{SubSecDQNbasedApproach}

In order to get an insight into the benefits of the above kernel-based approach, this subsection elaborates on a baseline, where state-of-the-art fully-connected DNNs \cite{GLSLZ23} are employed to approximate the action-values of $Q_{e,m}(\mathbf{s}_{t},\mathbf{a}_{t,m};\mathbf{w}_{e,m})$ and $Q_{d,m}(\mathbf{s}_{t},\mathbf{a}_{t,m};\mathbf{w}_{d,m})$.
Adam optimizer \cite{mnih2015human} and experience replay are exploited to optimize $\mathbf{w}_{e,m}$ and $\mathbf{w}_{d,m}$.
%The replay memories are denoted as $B_{e,m}$ and $B_{d,m}$, respectively.
Thus, in each iteration of the proposed algorithm, a minibatch of $N$ transition samples $(\mathbf{s}_t, \mathbf{a}_{t,m}, \mathbf{r}_{m,t}, \mathbf{s}_{t+1})$ is randomly taken from $B_{e,m}$ as well as $B_{d,m}$ for optimizing $\mathbf{w}_{e,m}$ and $\mathbf{w}_{d,m}$. The optimizations can be respectively formulated as
%\begingroup\makeatletter\def\f@size{9}\check@mathfonts
%\def\maketag@@@#1{\hbox{\m@th\small\normalfont#1}}%
%\begin{IEEEeqnarray}{rcl}
%\begin{split}
%\mathbf{w}_{e,m}\!\leftarrow\!\textstyle{ \arg \min_{{\mathbf{w}_{e,m}}}}\tfrac{1}{N}\textstyle{\sum_{{k=1}}^{N}}{\left\vert y_{e,m,k}\!-\! Q_{e,m}(\tilde{\mathbf{s}}_{k},\mathbf{a}_{m,k};\mathbf{w}_{e,m}) \right\vert}^{2}, \\
%\mathbf{w}_{d,m}\!\leftarrow\!\textstyle{ \arg \min_{{\mathbf{w}_{d,m}}}}\tfrac {1} {N}\textstyle{\sum_{k=1}^{N}}{\left\vert y_{d,m,k}\!-\!Q_{d,m}(\tilde{\mathbf{s}}_{k},\mathbf{a}_{m,k};\mathbf{w}_{d,m}) \right\vert}^{2},
%\end{split}
%\nonumber
%\end{IEEEeqnarray}
%\endgroup
\begin{IEEEeqnarray}{rl}
\mathbf{w}_{e,m}{}\!\leftarrow\! \arg \min_{{\mathbf{w}_{e,m}}}\frac{1}{N}{\sum_{{k=1}}^{N}}\left| y_{e,m,k}\!-\! Q_{e,m}\!(\tilde{\mathbf{s}}_{k},\mathbf{a}_{m,k};\mathbf{w}_{e,m})\! \right|^2
\end{IEEEeqnarray}
and
\begin{IEEEeqnarray}{rcl}
\mathbf{w}_{d,m}\!\leftarrow\!{ \arg \min_{{\mathbf{w}_{d,m}}}}\frac {1} {N}{\sum_{k=1}^{N}} \left| y_{d,m,k}\!-\!Q_{d,m}\!(\tilde{\mathbf{s}}_{k},\mathbf{a}_{m,k};\mathbf{w}_{d,m}\!) \right|^2,
\end{IEEEeqnarray}
where $y_{e,m,k}\!=\!e_{k+1}+\gamma_r\textstyle{\max_{\mathbf{a}_{m}}}Q_{e,m}(\tilde{\mathbf{s}}_{k+1},\mathbf{a}_{m};\mathbf{w}_{e,m}^{-})$ and $y_{d,m,k}\!=\!d_{k+1}+\gamma_r \textstyle{\max_{\mathbf{a}_{m}}}Q_{d,m}(\tilde{\mathbf{s}}_{k+1},\mathbf{a}_{m};\mathbf{w}_{d,m}^{-})$ represent the temporal difference targets, and the target network weights $\mathbf{w}_{e,m}^{-}$ and $\mathbf{w}_{d,m}^{-}$ can be iteratively updated with the optimized $\mathbf{w}_{e,m}$ and $\mathbf{w}_{d,m}$ \cite{1998Reinforcement}.

\section{Numerical Results}
\label{SecSimResults}
\begin{figure}[t]
\centering
\includegraphics[width = 3.4in]{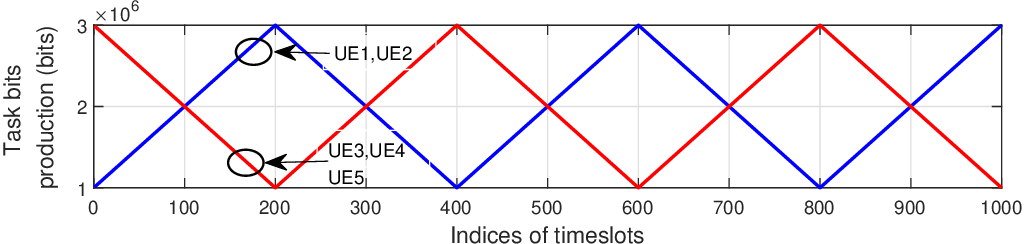}
\caption{Number of task bits produced in each UE over timeslots.}
\label{taskproduction.eps}
\end{figure}

In the simulations, $\mathcal{A}_0$ contains eight cardinal directions and $H = 100$\,m.
Regarding the channels, the small-scale fading is characterized by Rayleigh fading and the pathloss at a reference distance of $1$m is set as $39$\,dB. The LoS probability $P_\text{LoS}$ is characterized by the constant modeling parameters  $a=9.61$ and $b=0.16$ \cite{SD2022}. Additionally, the pathloss exponent is set as $\beta = 2.6$. For the UEs, $M = 5$, $P_m =$ 30\,dBm, $\sigma_n^2 = -90$dBm and $B = 6$\,MHz \cite{Wang2022DynamicAC}.
For task processing, $\tau = 2$\,s; $c =$ 1e3\,cycles/bit \cite{HWYZ19};
$f_{\text{UAV}} =$ 1.6e9\,cycles/s, $f_{\text{BS}} =$ 1.8e9\,cycles/s and $f_{\text{UE}} =$ 8e8\,cycles/s;
$\kappa_\text{UE} =$ 1e-28, $\kappa_\text{BS}$ = 1e-28 and $\kappa_\text{UAV}$ = 1e-27.
The thresholds for recognizing a new state are set as $\mu_q=2$ and $\mu_d=0.3$.
For both kernel-based and DNN-based approaches, $w_e = 1$, $w_d = 1$ and $\gamma_r = 0.3$, unless otherwise stated.
In the kernel-based approach, $\sigma_{s_1} = 200$, $\sigma_{s_2} = 1$ and $\sigma_a = 1$; $\mu_0 = 0.82$ and $n = 5$ (for the $n$-step return), unless otherwise stated.
In the DNN-based approach, each action-value is approximated by a fully-connected DNN with 3 hidden layers, where each layer consists of 64 neurons and tan-sigmoid is selected as the activation function; the size of the minibatch is set as $N=64$.
The simulations are conducted by MATLAB R2020a on a single computer, with an Intel Core i7 processor at 3.6GHz, a RAM of 16GB and the Windows 10 operating system.
In order to gain insights into the proposed approach, we consider that the number of task bits produced at each UE is periodic, as shown in Fig. \ref{taskproduction.eps}.

\begin{figure}[t]
\centering
\includegraphics[width = 3.5in]{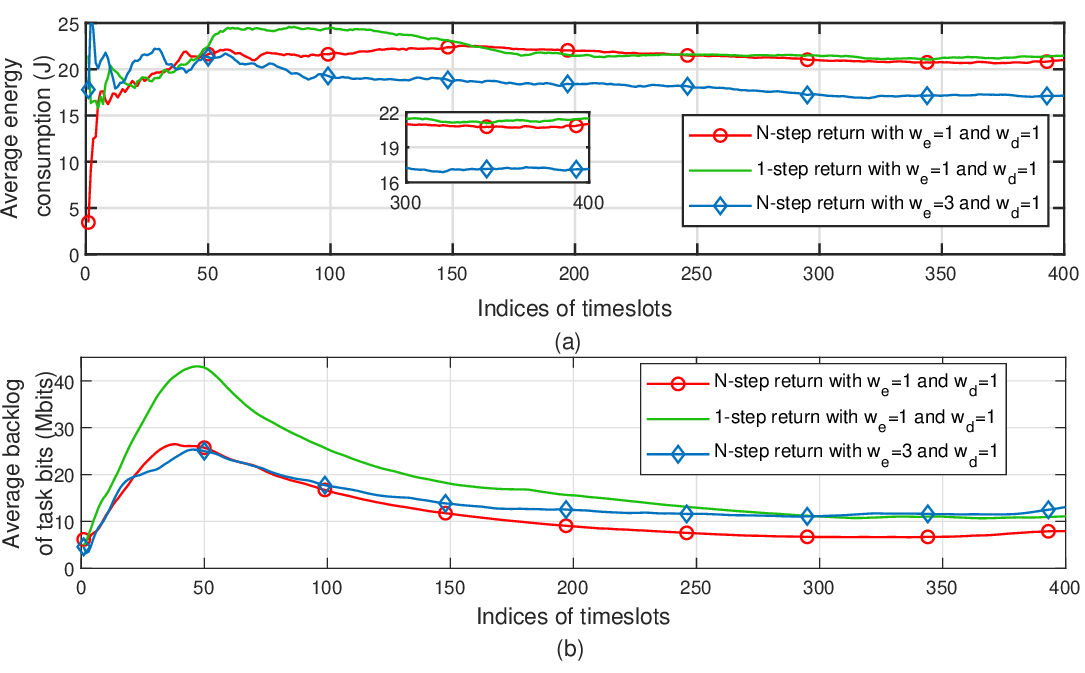}
\caption{Performance achieved by kernel-based approaches with $n$-step return (for $n=30$) at different weights and 1-step return. (a) Average energy consumption as a function of timeslots. (b) Average backlog of task bits as a function of timeslots.}
\label{duibi.eps}
\end{figure}
Fig. \ref{duibi.eps} investigates the long-term average energy consumption and the long-term average backlog performance achieved by kernel-based approaches with $n$-step return at different weights and 1-step return. It can be seen that both the average energy consumption and the average backlog achieved by the kernel-based approaches finally converges. Since the algorithms perform online learning, the average backlog of task bits first increases with the increasing time. Then, as the proposed algorithms continuously optimize the trajectory planning and offloading scheduling policies, the average backlog of task bits decreases. Fig. \ref{duibi.eps} also illustrates that since the $n$-step return can average the fluctuates of immediate rewards especially $d_t$, the average backlog of task bits achieved by the approach with $n$-step return can be significantly lower than that achieved by the approach with $1$-step return.
Moreover, Fig. \ref{duibi.eps} also indicates that in the presence of $w_e = 3$ and $w_d = 1$, the average energy consumption achieved by the kernel-based approach with $n$-step return can yield lower energy consumption but higher average backlog of task bits than that with $w_e = 1$ and $w_d = 1$. The reason lies in that a higher $w_e$ can induce the dominance of maximizing $Q_{e,m}(\tilde{\mathbf{s}}_{t}, \mathbf{a}_{t,m})$ in the objective function of problem (\ref{Select_action}), leading to an decrease in the average energy consumption. The formulation of $E_t$ indicates that the actions $\mathbf{a}_{t,m} = [\alpha_{m,t,\text{UAV}}, \alpha_{m,t,\text{BS}}, \alpha_{m,t,\text{UE}}]^T$ are related to not only energy consumption but also offloading/computing. Therefore, an decrease in energy consumption can cause an increase in the average backlog of task bits.

\begin{table}[t]
\centering
\caption{Average Decision-Making \& Online Learning Time.}
\label{TableAvgDecisionTime}
\begin{tabular}{|c|c|c|}
\hline
    Algorithms & DNN-based & Kernel-based\\ [0.3ex]
\hline
    \makecell{Elapsed time (s)} & 1.511 & 0.0045 \\ [0.3ex]
\hline
\end{tabular}
\end{table}

\begin{figure}[t]
\centering
\includegraphics[width = 3.6in]{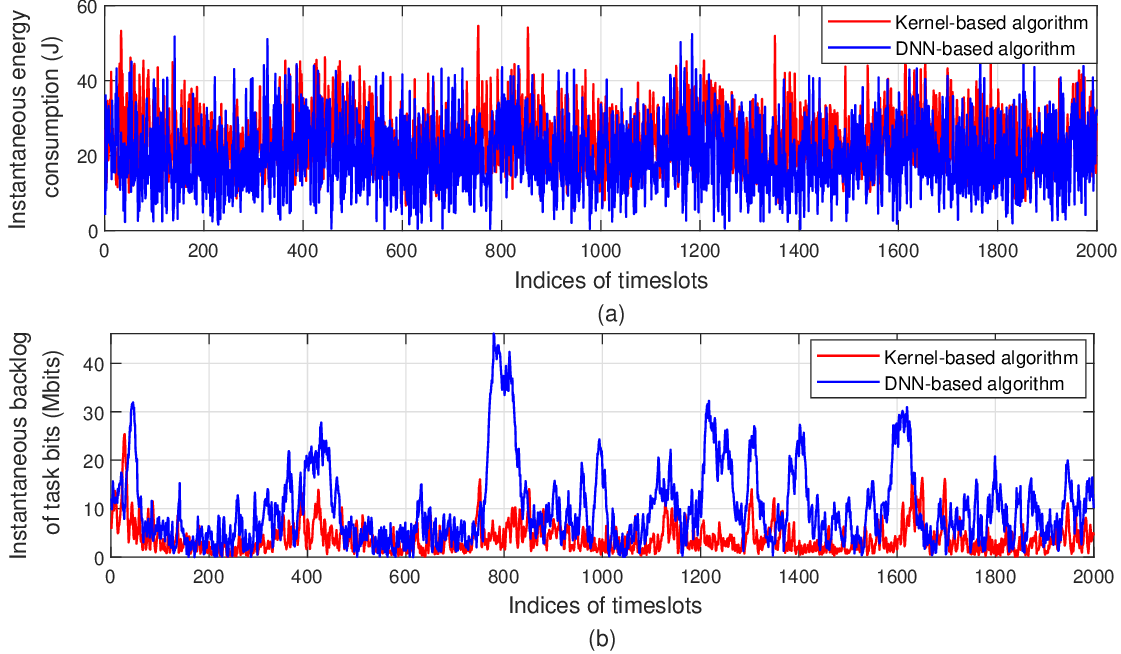}
\caption{Performance comparison of the kernel-based approach and the DNN-based approach after running for 7000 timeslots. (a) Instantaneous energy consumption as a function of timeslots. (b) Instantaneous backlog of task bits as a function of timeslots.}
\label{DNNandkernel.eps}
\end{figure}
Table \ref{TableAvgDecisionTime} depicts the average elapsed time of decision making and online learning in each timeslot achieved by the kernel-based and the DNN-based approach.
It is shown that the kernel-based approach consumes significantly less time than the DNN-based approach.
Meanwhile,
Fig. \ref{DNNandkernel.eps} indicates the performance achieved by the kernel-based approach and the DNN-based approach during timeslot 7001 and timeslot 9000.
By performing online learning for a duration of 7000 timeslots, it can be inferred from the numerical results as shown in Fig. \ref{duibi.eps} that the long-term average rewards achieved by the algorithms converge.
It can be observed from Fig. \ref{DNNandkernel.eps} that although the network performing the kernel-based approach achieves slightly lower energy consumption than that performing the DNN-based approach, the latter suffers from dramatic fluctuations in the instantaneous backlog of task bits.
This is due to that the kernel-based approach can benefit from the design of $n$-step return and the neural networks consisting of kernel functions.
The sizes of such neural networks can be adaptive to the environments by adding more appropriate decision-making features, yielding more accurate approximation of actor-values.
Furthermore, since the number of task bits produced by the cluster of UEs 3, 4 and 5 reaches peaks in timeslots 400, 800, 1200, etc. (as depicted in Fig. \ref{taskproduction.eps}),
the instantaneous backlog of task bits achieved by the kernel-based approach around timeslots 400, 800, 1200, 1600 and 2000 is sightly higher than that in other timeslots.

\begin{figure}[t]
\centering
\includegraphics[width = 3.5in]{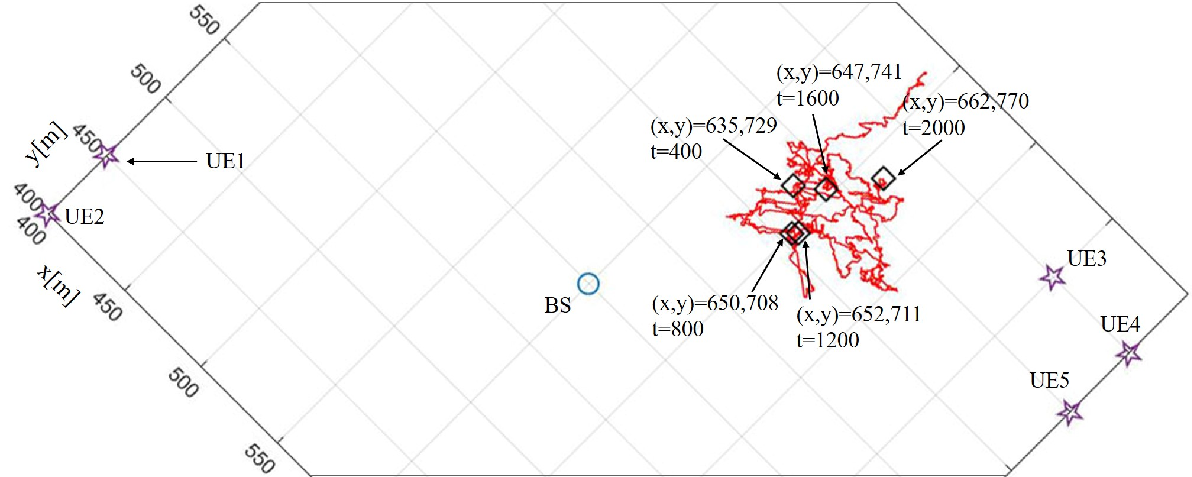}
\caption{The UAV's trajectory, where the notations $(x, y)$ and $t$ represent the horizontal position of the UAV and the index of a timeslot, respectively.}
\label{UAVguiji.eps}
\end{figure}
Fig. \ref{UAVguiji.eps} illustrates the UAV's trajectory achieved by the kernel-based approach during the duration of the 2000 timeslots as shown in Fig. \ref{DNNandkernel.eps}.
As the average number of task bits per timeslot produced by the cluster of UEs 3, 4 and 5 is greater than that produced by the cluster of UEs 1 and 2 (as shown in Fig. \ref{taskproduction.eps}),
the UAV always hovers at the right hand side of the BS, such that the overall network can benefit from the stronger air-ground channel and the edge server at the UAV.

\section{Conclusions}
\label{SecConclusions}
We have proposed a novel multi-objective trajectory planning and offloading scheduling scheme based on RL for dynamic air-ground collaborative MEC. In order to address the issues of multi-objective MDP and the curses of dimensionality caused by multiple UEs, the scheme is developed based on a distributed structure, where MORL and the kernel method are integrated. Numerical results reveal that benefiting from the design of $n$-step return, the proposed approach can outperform the design with 1-step return. Moreover, due to the $n$-step return and the kernel-based neural networks, the proposed kernel-based approach can significantly outperform the DNN-based approach in terms of the backlog of task bits and the average decision-making and online learning time.

%Simulation results show that this strategy enables joint trajectory planning and offloading scheduling to adapt to dynamic computational demands.
%\footnotesize
\bibliographystyle{IEEEtran}
\bibliography{IEEEabrv,reference}
\end{document}